\documentstyle[12pt]{article}
\def\lapproxeq{\lower .7ex\hbox{$\;\stackrel{\textstyle <}{\sim}\;$}}
\def\gapproxeq{\lower .7ex\hbox{$\;\stackrel{\textstyle >}{\sim}\;$}}

\begin{document}

\titlepage

\baselineskip12pt
\begin{flushright}
\begin{tabular}{l}
{\bf hep-ph/9403246}\\
March 1994
\end{tabular}
\end{flushright}

\vspace*{1cm}
\begin{center}
{\bf Top quark polarization as a probe of $t\bar{t}$ threshold dynamics}
\end{center}
\vspace*{1.5cm}

\begin{center} V.S.\ Fadin \\ Budker Institute for Nuclear Physics \\ and \\
Novosibirsk State University, 630090 Novosibirsk, Russia.
\end{center}

\begin{center}
V.A.\ Khoze \\ Department of Physics, University of Durham, DH1 3LE, England
\\
and \\
Institute for Nuclear Physics, 188350 St.\ Petersburg, Gatchina, Russia.
\end{center}

\begin{center}
M.I.\ Kotsky \\ Budker Institute for Nuclear Physics, 630090 Novosibirsk,
Russia.
\end{center}

\vspace*{1.5cm}

\begin{abstract}
We study the spin properties of top quarks produced in collisions of polarized
photons in the threshold region.  For a relatively heavy top quark the
influence of non-perturbative effects is small and its polarization parameters
can be predicted in perturbative QCD.  The measurements of the top
polarization may allow a novel test of QCD in the $t\bar{t}$ system.  In
particular, they may provide a new way to determine the precise value of
$\alpha_s$ and to study the properties of the top quark.
\end{abstract}

\newpage

\noindent 1.  INTRODUCTION \\
\indent The top quark has not yet been discovered and the Fermilab experiments
have steadily increased its mass limit (which is now well above 100
GeV)$^{[1]}$.  The indirect evidence for the existence of top quark is very
strong and if it behaves as predicted by the minimal Standard Model then its
mass should be in the range of 100-200 GeV or thereabout, e..g.\ [2,3].  This
range suggests that the $t$-quark has a good chance to be detected in the
foreseeable future.

Since the top mass $m_t$ exceeds the $W$-mass, the $t$-quark can decay
directly to $W^+b$ and its decay width $\Gamma_t$ is steeply increasing with
its mass$^{[4,5]}$.  For $m_t \sim$ 130-150 GeV the $t$-quark width is
typically in the GeV range$^{[6]}$ and the $t$ generally decays before its
hadronization occurs$^{[4,5]}$.  In such a situation, the $t$-quark spin
proves to be a very useful tool.  The point is that the $t$-polarization is
not diluted since there is no hadronization and the bremsstrahlung gluons do
not flip the top spin.  Therefore the $t(\bar{t})$ quark is normally produced
in a well-defined state of polarization which can be measured experimentally.
This polarization    is transmitted to the secondary particles produced in
their decays and can be well reconstructed from their distributions (see e.g.\
Refs.\ [7-11]).

Recall that the top-quark polarization is just one source of the
$P$-odd angular asymmetry of the secondary particles induced by the weak top
decay.  Some other parity-violating effects in the final state distributions
could be of practical interest.

A number of important issues of top quark dynamics including the polarization
phenomena could be addressed at the next-generation machine, a linear $e^+e^-$
collider with centre-of-mass energy in a first phase between 300-500
GeV$^{[9-10]}$.  One of the main aims here is the detailed study of the
properties of $t\bar{t}$ production near threshold in $e^+e^-$ annihilation
\begin{equation}
e^+e^- \rightarrow t\bar{t}.
\end{equation}
The new opportunities would be provided by a Photon Linear Collider (PLC).
This is
a facility$^{[12-14]}$ where high energy, high intensity photon beams are
generated at a linear $e^+e^-$ collider via Compton backscattering using
high-power lasers$^{[15]}$.  PLC provides highly
polarized beams, large luminosity, and a variable luminosity spectrum.  The
potential of such a machine to explore top quark threshold production in
\begin{equation}
\gamma\gamma \rightarrow t\bar{t}
\end{equation}
was discussed in Refs.\ [16-18].

Since the cross-section of reaction (2) in the threshold region is large
enough, it may be possible to perform here some new tests of the properties of
the top quarks.  Recall that (predominantly $S$-wave) $t\bar{t}$ threshold
cross section for process (1) can be described in terms of the Green's
function $G_E(\vec{r} = 0, \bar{r}^{\prime} = 0)$ of the Schr\"{o}dinger
equation for the interacting $t\bar{t}$ system$^{[19-21]}$.  Collisions of
polarized photons off each other allow one to separate the $P$-wave production
and, thus, to study experimentally the second space derivative of this Green's
function$^{[17,22]}$.

Measurements of the top polarization in the process (2) may provide a novel
probe of QCD dynamics in the $t\bar{t}$ system.  Here one deals with the
interference between the $S$-wave and $P$-wave production.  These measurements
may provide information not only about the overlap of $S$-wave and $P$-wave
states of the $t\bar{t}$ system but also about the phase difference of the
corresponding production amplitudes.

In terms of the coordinate Green's function for the $t\bar{t}$ system one
studies here the Fourier transform of the space derivative
$\frac{\partial}{\partial \vec{r}^{\prime}} G_E(\vec{r},\vec{r}^{\prime} = 0)$
of this function.  This provides an opportunity to perform important studies
of the behaviour of the QCD potential especially at large distances.  Another
way to probe the effects of the $S-P$ interference by measuring the
forward-backward asymmetry $\Delta_{FB}$ in the process (1) was advocated in
Ref.\ [23].

The remainder of the paper is organized as follows.  In section 2 we consider
the spin properties of the top quarks produced in collisions of polarized
photons.  In section 3 we describe the formalism needed for incorporating the
final state interaction effects.  In section 4 the threshold expressions for
the polarization parameters are summarized.  Some discussion is presented in
section 5.  Section 6 gives our conclusions.

\vspace*{1cm}
\noindent 2.  POLARIZATION PHENOMENA IN TOP PRODUCTION IN
$\gamma\gamma$-COLLISIONS \\
\indent The $t\bar{t}$ production cross section in photon-photon fusion (2)
depends sensitively on the initial photon polarizations and on the top
spin-vector $\vec{\zeta}^{\prime}$.  For the purposes of illustration it is
useful to
first gain insight by studying the polarization effects appearing in Born
approximation, neglecting for a moment the width of the top.

We begin by defining some variables: $\sqrt{s}$ is the total centre-of-mass
energy, $E = \sqrt{s}-2m_t$ is the energy excess above the threshold and
$\beta = \sqrt{1 - \frac{4m_t^2}{S}}$ is the velocity of the on-shell
$t$-quarks.  $\vec{k}_1 = \frac{\sqrt{s}\vec{n}_1}{2}$ and
$\vec{k}_2 = - \frac{\sqrt{s}\vec{n}_1}{2}$ are the momenta of the
incoming photons in the lab ($\gamma\gamma$ centre-of-mass) frame and $\vec{p}
= p \vec{n}_2$ is the $t$-quark momentum, $\theta$ is the polar angle of the
$t$ with respect to $\vec{n}_1$.  We introduce the Stokes parameters
$\xi^{(1)}_i, \, \xi^{(2)}_i$ of the initial photons referred to orthonormal
vectors related to the production plane\footnote{In Refs.\ [17,24] a different
orientation of these axes was used.}
\begin{displaymath}
\vec{\chi}^{(1)}_2 \; = \; \vec{\chi}^{(2)}_2 \; = \; \frac{(\vec{n}_1 \times
\vec{n}_2)}{|(\vec{n}_1 \times \vec{n}_2)|} ,
\end{displaymath}

\begin{equation}
\vec{\chi}^{(1)}_1 \; = \; -\vec{\chi}^{(2)}_1 \; = \; (\vec{\chi}^{(1)}_2
\times \vec{n}_1) .
\end{equation}

The general expression$^{[24]}$ for the differential Born cross-section for
production of the $t$-quark with polarization $\vec{\zeta}^{\prime}$ in the
collisions of the arbitrary polarized photons is rather complex.  In the
threshold region ($\beta \ll 1$) we may neglect the order-$\beta^2$
corrections and the non-relativistic Born cross section is given by a simple
formula
$$
\frac{d\sigma_{{\rm nonrel}}^{(B)}}{d{\rm cos}\theta d\phi} \;  = \;
\frac{\alpha^2R_{\gamma\gamma}}{2s} \beta \left\{ 1 + \xi^{(1)}_1 \xi^{(2)}_1
+ \xi^{(1)}_2 \xi^{(2)}_2 - \xi^{(1)}_3 \xi^{(2)}_3  \right.
$$

$$
+ 2 \beta \left[ (\xi^{(1)}_2 + \xi^{(2)}_2)(\vec{\zeta}^{\prime}\cdot
\vec{n}_2) - (\xi^{(1)}_2 \xi^{(2)}_3 + \xi^{(1)}_3 \xi^{(2)}_2)
((\vec{\zeta}^{\prime}\cdot \vec{n}_1) {\rm cos} \theta -
(\vec{\zeta}^{\prime}\cdot \vec{n}_2))  \right.
$$

\begin{equation}
\left. \left. - (\xi^{(1)}_2 \xi^{(2)}_1 - \xi^{(1)}_1 \xi^{(2)}_2)\cdot
\vec{\zeta}^{\prime}\cdot (\vec{n}_1 \times \vec{n}_2) \right] \right\}
\end{equation}
with
\begin{equation}
R_{\gamma\gamma} \; = \; N_c\cdot e_t^4
\end{equation}
where the number of colours is denoted by $N_c = 3$ and the electric charge of
top-quark by $e_t = \frac{2}{3}$.

One can easily see from Eq.\ (4) that the $t$-quark polarization
$\vec{\zeta}^{\prime}$ enters only in correlation with a circular polarization
of a photon $\xi^{(1)}_2$ or $\xi^{(2)}_2$.  Therefore in Born approximation
linearly polarized photons cannot generate polarization of the top as it
follows from unitarity and T-invariance.
A top-quark polarization perpendicular to the production plane occurs only in
combination with the parameter
\begin{equation}
C^{(-)}_{21} \; = \; (\xi^{(1)}_2 \xi^{(2)}_1 - \xi^{(1)}_1 \xi^{(2)}_2)
\end{equation}
and the projection on the production plane enters the expression for the cross
section only with the parameters
\begin{equation}
C^{(+)}_2 \; = \; (\xi^{(1)}_2 + \xi^{(2)}_2), \hspace*{1cm} C^{(+)}_{23} \; =
\; (\xi^{(1)}_2 \xi^{(2)}_3 + \xi^{(1)}_3 \xi^{(2)}_2) .
\end{equation}
These observations are the results of the identity of photons and of the $P$
and $T$ invariances in the lowest order in perturbation theory.  They remain
valid at photon energies far above the $t\bar{t}$ production
threshold$^{[24]}$.

For the purposes of the forthcoming discussion of the effects generated by the
QCD Coulomb forces and by the $t$-quark instability we present here an
expression for the nonrelativistic Born amplitude $A^{(B)}_{fi}$ for process
(2).  Neglecting contributions of relative order of $O\left( \frac{p^2}{m_t^2}
\right)$ this amplitude can be written in terms of the two-component wave
functions   $\psi(\chi)$ describing the spin states of the produced top quark
(antiquark) as
\begin{equation}
A^{(B)}_{fi}(\vec{p}) \; = \; 4\pi\alpha e^2_t \psi^+ \left( a + \frac{p}{m_t}
\vec{b}\cdot \vec{\sigma}\right)\chi
\end{equation}
where the quantities $a$ and $b_i$ are related to the photon polarization
vectors $\vec{e}^{(1)}, \vec{e}^{(2)}$ by
\begin{displaymath}
a \; = \; 2i (\vec{e}^{(1)} \times \vec{e}^{(2)}) \vec{n}_1 ,
\end{displaymath}
\begin{equation}
\vec{b} \; = \; 2[(\vec{e}^{(1)}\cdot \vec{e}^{(2)})\vec{n}_1 {\rm cos}\theta
+ (\vec{e}^{(1)}\cdot \vec{n}_2) \vec{e}^{(2)} + (\vec{e}^{(2)}\cdot
\vec{n}_2) \vec{e}^{(1)}] .
\end{equation}

The first (second) term in the brackets in (8) corresponds to $S$-wave
($P$-wave) states of the $t\bar{t}$ pair.  Thus the top quarks are produced
near the threshold of the process (2) in $S$-wave and $P$-wave states with
$d\sigma^{(B)}_S \sim \beta$ and $d\sigma^{(B)}_P \sim \beta^3$ respectively.
The $S$-wave and $P$-wave production cross section can be separated in the
collisions of circularly polarized photons, e.g. [17].  If the two photons
carry opposite helicities $(\xi^{(1)}_2 = -\xi^{(1)}_2 = \pm 1)$ the
$t\bar{t}$ pair is produced in a pure $P$-wave state.  It is the interference
of $S$-wave and $P$-wave amplitudes which gives rise to the $t$-quark
polarization at order $\beta$.

It is quite straightforward to demonstrate that
\begin{equation}
|a|^2 \; = \; 2(1 + C^{(0)})
\end{equation}
with
\begin{equation}
C^{(0)} \; = \; (\xi^{(1)}_1 \xi^{(2)}_1 + \xi^{(1)}_2 \xi^{(2)}_2 -
\xi^{(1)}_3 \xi^{(2)}_3)
\end{equation}
and
\begin{equation}
Re(a^*\vec{b}\cdot\vec{\zeta}^{\prime}) \; = \; 2\left[
C^{(+)}_2(\vec{\zeta}^{\prime}\cdot\vec{n}_2) -
C^{(+)}_{23}((\vec{\zeta}^{\prime}\cdot\vec{n}_1){\rm cos}\theta -
(\vec{\zeta}^{\prime}\cdot\vec{n}_2)) - C^{(-)}_{21} \vec{\zeta}^{\prime}
(\vec{n}_1 \times \vec{n}_2) \right] .
\end{equation}
Summing $|A^{(B)}_{fi}|^2$ over colours and the antitop polarizations and
making use of Eqs.\ (10-12) one can easily recognize Eq.\ (4) as
\begin{equation}
\frac{d\sigma^{(B)}_{{\rm nonrel}}}{d{\rm cos}\theta d\phi} \; = \;
\frac{\alpha^2R_{\gamma\gamma}}{4s}\cdot \beta \left[|a|^2 + \frac{2p}{m_t}
Re(a^*\vec{b}\cdot \vec{\zeta}^{\prime})\right].
\end{equation}
When the final state QCD interaction is switched on, the interference between
$S$-wave and $P$-wave production amplitudes becomes complex and
$Im(a^*\vec{b}\vec{\zeta}^{\prime})$ also may enter the expression for the
cross section.  It is given by
\begin{displaymath}
Im(a^*\vec{b}\cdot\vec{\zeta}^{\prime}) \; = \; -2 \left[
C^{(+)}_1((\vec{\zeta}^{\prime}\cdot\vec{n}_1){\rm cos}\theta -
(\vec{\zeta}^{\prime}\cdot \vec{n}_2)) \right.
\end{displaymath}

\begin{equation}
\left. - C^{(+)}_{13}(\vec{\zeta}^{\prime}\cdot \vec{n}_2) + C^{(-)}_3
\vec{\zeta}^{\prime}\cdot (\vec{n}_1 \times \vec{n}_2) \right]
\end{equation}
with
\begin{displaymath}
C^{(+)}_1 \; = \; (\xi^{(1)}_1 + \xi^{(2)}_1) , \hspace*{1cm} C^{(-)}_3 \; =
\; (\xi^{(1)}_3 - \xi^{(2)}_3) ,
\end{displaymath}

\begin{equation}
C^{(+)}_{13} \; = \; (\xi^{(1)}_1 \xi^{(2)}_3 + \xi^{(1)}_3 \xi^{(2)}_1) .
\end{equation}

Note that these ($T$-odd) correlations between the spin-vector
$\vec{\zeta}^{\prime}$ and the photon polarization parameters $C^{(+)}_1,
C^{(-)}_3, C^{(+)}_{13}$ are the only ones that survive the restrictions
imposed by parity conservation and the identity of photons.  Such correlations
cannot appear in the circularly polarized photon beams.

We return now to the polarization parameters of the $t$ in Born approximation.
They can be easily derived from Eqs.\ (4),(13).  Thus,
\begin{equation}
\vec{\zeta}_B \; = \; 2\beta \cdot Re \vec{\zeta}_0 , \hspace*{1cm}
\vec{\zeta}_0 \; = \; \frac{a^*\vec{b}}{|a|^2} .
\end{equation}

Let us choose a coordinate system where $\vec{p}$ is along the z-axis and
$\vec{p} \times \vec{k}_1$ along the y-axis.  Then the $t$-quark helicity $P_L
= \zeta_z$, the transverse polarization within the production plane $P_{\bot}
= \zeta_x$, and the spin component normal to this plane $P_N = \zeta_y$, are
\begin{displaymath}
P_L \; = \; \frac{2\beta}{1 + C^{(0)}} (C^{(+)}_2 + {\rm sin}^2\theta
C^{(+)}_{23}) ,
\end{displaymath}

\begin{displaymath}
P_{\bot} \; = \; -\frac{\beta}{1 + C^{(0)}} C^{(+)}_{23}\cdot {\rm sin}2
\theta,
\end{displaymath}

\begin{equation}
P_N \; = \; \frac{2\beta C^{(-)}_{21}}{1 + C^{(0)}}\cdot {\rm sin}\theta,
\end{equation}
with the photon polarization parameters $C$ given by Eqs.\ (6,7,11).

Note that if the Stokes parameters are defined in a reference frame where the
vector $\vec{n}_2$ is not related to the top momentum, the expressions for the
parameters $C^{(0)}$ and $C^{(+)}_2$ are valid for any choice of $\vec{n}_2$
(except for $\vec{n}_2 \parallel \vec{n}_1$).  On the contrary, if the
orientation of $\vec{n}_2$ is fixed, the parameters $C^{(-)}_{21}$ and
$C^{(+)}_{23}$ vanish after integration over the directions of the top
momentum.  As a result, the top-quark helicity averaged over the angles is
given by
\begin{equation}
\langle P_L\rangle \; = \; \frac{2\beta (\xi^{(1)}_2 + \xi^{(2)}_2)}{1 +
C^{(0)}} .
\end{equation}
It is worth mentioning here that the top-quark polarization perpendicular to
the production plane arises even at the Born level.  This is in contrast to
the process (1) where (assuming CP conservation) the normal polarization can
be induced only by the the higher loop corrections.

Close to the production threshold the Born result for the top polarization is
suppressed by the $P$-wave factor $\beta$.  Away from the threshold this
polarization is quite sizeable.  For the highly polarized photon
beams $|\vec{\zeta}_0| \sim O(1)$.  For instance, if two photons carry the
same helicities ($\xi^{(1)}_2\cdot \xi^{(2)}_2 = +1$) then $(\zeta_0)_L =
|\vec{\zeta}_0| = 1$.  Moreover, only one photon can be polarized to generate
the $t$-quark polarization.  Recall that the evident advantage of the
polarization measurements is that they are not sensitive to the normalization
of the cross-section.  Finally, let us note that in the case when two photons
carry the opposite helicities ($\xi^{(1)}_2\cdot \xi^{(2)}_2 = -1$) the
formulae presented here cannot be applied since we systematically neglected
$O(\beta^3)$ contributions.

Up to now we have not taken into account the width of the top, which should be
included in our analysis for a correct discription of the threshold
production.  In Born approximation this changes only the phase space factor.
Taking into account that in the
phase space integral for an unstable quark the mass-shell factor
$\delta(p^2-m^2_t)$ should be replaced by
\begin{equation}
\Delta (p^2) \; = \; \frac{m_t\Gamma_t}{\pi[(p^2-m^2_t)^2 + m^2_t\Gamma_t^2]}
,
\end{equation}
one can approximate the two-particle phase space element $d\rho$ in the
non-relativistic limit by
\begin{equation}
d\rho \; = \; \frac{d^3p}{(2\pi)^3} \frac{2\Gamma_t}{(2m_t)^2 \left[ \left( E
- \frac{p^2}{m_t} \right)^2 + \Gamma_t^2 \right]} .
\end{equation}

Therefore, the width effects in Born approximation for the threshold
cross-section lead to the replacement of $\beta d{\rm cos}\theta d\phi$ in
Eqs. (4),
(13) by $8(2\pi)^2d\rho$, where $d\rho$ is given by Eq. (20).

\vspace*{1cm}
\noindent 3.  EFFECTS OF THE FINAL STATE COULOMB INTERACTION \\
\indent The striking features of the $t\bar{t}$ production in the threshold
region have been discussed quite frequently in recent years (e.g.\ [9-11] and
references therein).  One faces here the unique situation that the QCD binding
forces are strong but (because of the infrared cut-off provided by the large
top width) are well under the control of perturbation theory$^{[19-21]}$.

The $t\bar{t}$ production near threshold is significantly enhanced because of
the Coulomb gluon exchanges between the top quarks.  To take full advantage of
the polarization measurements one needs to understand the following questions.
First, how strongly the top polarization is affected by the Coulomb threshold
interaction.  Second, how the top polarization can be exploited to obtain some
new information on the $t\bar{t}$ interaction forces.  Answering these
questions is the main concern of this and the next section.

In what follows we shall use a non-relativistic Green's function formalism
applied in the previous papers of two of the authors for finding the total
cross-section$^{[19-20]}$ and the amplitude$^{[20]}$ for process (1), see also
Refs.\ [21-23,25,26].  As in Refs.\ [19-20] we shall use the Coulomb gauge
keeping only the instantaneous Coulomb part of the gluon propagator,
\begin{displaymath}
D^{\mu \nu}(q^2) \; = \; i \frac{\delta^{\mu o}\delta^{\nu o}}{\vec{q}^2} .
\end{displaymath}
Since the interaction vertex of the transverse gluons with the quarks is
proportional to the relative quark velocity, $v$,
\begin{equation}
v \; = \; \frac{2p}{m_t}
\end{equation}
($v = 2\beta$ on the mass-shell) such gluons can generate only $O(\alpha_s)$
corrections to the leading Coulomb result.

The exchanges of $n$ Coulombic gluons between the quark and antiquark in the
final state generate the order $\left( \frac{\alpha_s}{v}\right)^n$  terms
which can be summed up systematically in the Green's function approach.  The
effects of the top quark width $\Gamma_t$ are incorporated through the
replacement of the non-relativistic quark energy $E$ by $E + i\Gamma_t$ in the
Green's functions.  We should warn the reader that the non-relativistic
treatment of the unstable particle production in the $P$-wave case may have
some potential problems, see for details [22].  However, we shall show below
that these problems are quite manageable in the case of the top quark
polarization we are concerned with, see also Ref.\ [23].

The Coulomb interaction is spin-independent and if one keeps only the leading
(order $\left( \frac{\alpha_s}{v}\right)^n$) terms, then the multiple gluon
exchange cannot affect the spinor structure of the Born amplitude$^{[20]}$.
We shall demonstrate this explicitly for the threshold amplitude for the
process (2) and derive the Green's function representation for this amplitude.

Let us label the particle 4-momenta by $\gamma(k_1) + \gamma(k_2) \rightarrow
t(p_+) + \bar{t}(p_-)$ and present the amplitude in the centre-of-mass frame
in terms of the four-component quark wave functions $u(p_+)$ and $v(p_-)$ as
\begin{equation}
A_{fi}(\vec{p}) \; = \; 4\pi \alpha \cdot e_t^2 \bar{u}(p_+) M(\vec{p})
v(p_-).
\end{equation}
Up to order $\frac{p^2}{m_t^2}$ terms one can rewrite the quark spinors in a
form which simplifies the further treatment of the Dirac  structure of the
threshold amplitude
\begin{displaymath}
u(p_+) \; = \; \frac{(m_t + \hat{p}_+)}{\sqrt{2m_t}} \cdot \Lambda_+ u_0 ,
\end{displaymath}

\begin{equation}
v(p_-) \; = \; \frac{(m_t - \hat{p}_-)}{\sqrt{2m_t}} \cdot \Lambda_- v_0
\end{equation}
with
\begin{equation}
u_0 = \left(   \begin{array}{c}
\psi   \\   0
\end{array}  \right) ,  \hspace*{1cm} v_0 = \left(  \begin{array}{c}
0   \\   \chi
\end{array}   \right)    \hspace*{1cm}  \Lambda_{\pm} = \frac{(1 \pm
\gamma^0)}{2} .
\end{equation}
Then $A_{fi}$ takes the form
\begin{equation}
A_{fi} \; = \; 4\pi \alpha \, e^2_t u^+_0 R(\vec{p}) v_0
\end{equation}
where
\begin{equation}
R(\vec{p}) \; = \; \frac{1}{2m_t} \Lambda_+ \cdot (m_t + \hat{p}_+) M(\vec{p})
(m_t-\hat{p}_-) \cdot \Lambda_- .
\end{equation}
In Born approximation
\begin{displaymath}
M^{(B)}(\vec{p}) \; = \; \frac{\hat{e}^{(1)}(\hat{p}_+ - \hat{k}_1 +
m_t)\hat{e}^{(2)}}{2(k_1p_+)} +
\end{displaymath}

\begin{equation}
+ \frac{\hat{e}^{(2)} (\hat{p}_+ - \hat{k}_2 + m_t)\hat{e}^{(1)}}{2(k_2p_+)}
\; = \; e^{(1)}_{\mu} e^{(2)}_{\nu} T^{\mu\nu}_{(B)} (\vec{p})
\end{equation}
and $R(\vec{p})$ becomes
\begin{equation}
R^{(B)}(\vec{p}) \; = \; \Lambda_+ \left[ a \gamma^0 \gamma^5 + \frac{p}{m_t}
\vec{b}\cdot \vec{\gamma} \right] \Lambda_-,
\end{equation}

\begin{displaymath}
\gamma_5 \; = \; \left(  \begin{array}{cc}
O  &  I \\
I  & O      \end{array}   \right) ,
\end{displaymath}
cf.\ Eqs.\ (25-26) and (8).

Diagramatically, the final state Coulomb interaction is described by the sum
of the ladder diagrams shown schematically in Fig.\ 1, where $T^{\mu
\nu}_{(B)}(\vec{p})$ is given by Eq.\ (27).  Let us note that within our
approximation we can ignore the quark virtualities in the expression for
$T^{\mu \nu}_{(B)}$.

We need to sum these ladder diagrams keeping the contributions linear in the
quark momenta.  They can be induced only by the $\vec{\gamma}\cdot
\vec{p}^{(i)}$ terms in the numerators of the fermion propagators.  We recall
that in the nonrelativistic approximation the propagators of the quarks
(antiquarks) with the 4-momenta $p_+^{(i)} (p_-^{(i)})$ reduce to the form
\begin{equation}
\frac{i(m_t \pm \hat{p}^{(i)}_{\pm})}{(p^{(i)}_{\pm})^2 - m_t^2 +
im_t\Gamma_t} \rightarrow i\frac{\Lambda_{\pm} \mp
\frac{(\vec{p}_{\pm}^{(i)}\cdot\vec{\gamma})}{2m_t}}{\epsilon_{\pm}^{(i)} -
\frac{(\vec{p}^{(i)}_{\pm})^2}{2m_t} + i\frac{\Gamma_t}{2}} .
\end{equation}
Here $\epsilon_{\pm} = p_{\pm}^0 - m_t$ is the non-relativistic $t(\bar{t})$
energy.  Throughout this paper we shall take the width $\Gamma_t$ to be
constant evaluated on the mass-shell.

Examining the Dirac structure of a diagram of Fig.\ 1 easily shows that we
need to keep the $\vec{\gamma}\cdot\vec{p}^{(i)}$ terms only in the piece
\begin{equation}
K^{(B)}(\vec{p}^{(n)}) \; = \; (m_t + \hat{p}^{(n)}_{(+)}) M^{(B)}
(\vec{p}^{(n)}) (m_t - \hat{p}^{(n)}_{(-)}) .
\end{equation}
Indeed, since each Coulomb vertex generates only a $\gamma^0$ matrix, the
$\vec{\gamma}\cdot \vec{p}^{(i)}$ terms appearing in the other parts of a
ladder diagram are always sandwiched between the operators $\Lambda_+$ or
$\Lambda_-$ and, thus, cannot contribute.  If we compare Eq.\ (30) with Eqs.\
(23), (25) and (26) we can immediately conclude that the matrix structure of
$R^{(B)}(\vec{p})$ is not affected by the Coulomb interactions.  Then the
ladder diagrams can be summed up following the recipe suggested in Refs.\
[19,20] (see also [21,23,25,26]) according to which $R(\vec{p})$ should
satisfy the equation depicted diagramatically in Fig.\ 2.

We shall not discuss this recipe here in detail, but for the reader's
convenience we recall some of its key points.  First, since the Coulomb
propagator is independent of the gluon energy $q^0_i$ one can perform the
integration over $q^0_i$ by taking the residue (only one pole is located in
each half-$q^0_i$-plane).  Then we arrive at the non-relativistic perturbation
theory scenario for the $t\bar{t}$ system with the full Hamiltonian $\hat{H}$
\begin{equation}
\hat{H} \; = \; \hat{H}_0 + V(r), \hspace*{1cm} \hat{H}_0 \; = \;
\frac{\hat{\vec{p}}^2}{m_t}
\end{equation}
in the centre-of-mass frame.

Here $\hat{\vec{p}} = -i\frac{\partial}{\partial \vec{r}}$ is the momentum
operator and $V(r)$ is the nonrelativistic QCD potential for the $t\bar{t}$
system in the colour singlet state
\begin{equation}
V(r) \; = \; -\frac{4}{3} \frac{\alpha_s(r)}{r} .
\end{equation}
Analogously to [20] the solution to the equation illustrated by Fig.\ 2 can be
written as
\begin{equation}
R(\vec{p}) \; = \; \langle \vec{p}|(\hat{H} - E - i\Gamma_t)^{-1} R^{(B)}
(\hat{\vec{p}})|\vec{r} = 0\rangle \cdot \left( \frac{\vec{p}\,^2}{m_t} - E -
i\Gamma_t \right) ,
\end{equation}
where $|\vec{p} \rangle$ denotes a state of definite momentum $\vec{p} =
\vec{p}_+ = -\vec{p}_-$, and $|\vec{r}\rangle$ a state of definite relative
coordinate (we use normalization $\langle \vec{p}|\vec{r}\rangle =
e^{-i\vec{p}\cdot\vec{r}})$, $E = \sqrt{(p_+ + p_-)^2} - 2m_t$.

Let us define the functions $\tilde{G}(\vec{p},E)$ and
$\tilde{F}_i(\vec{p},E)$ by
\begin{displaymath}
\tilde{G}(\vec{p},E) \; = \; \langle \vec{p}|(\hat{H} - E -
i\Gamma_t)^{-1}|\vec{r} = 0 \rangle  \hspace*{1cm} (a)
\end{displaymath}

\begin{equation}
\tilde{F}_i(\vec{p},E) \; = \; \langle \vec{p}|(\hat{H} - E - i\Gamma_t)^{-1}
\hat{p}_i|\vec{r} = 0 \rangle  \hspace*{.75cm} (b)
\end{equation}

Note that to avoid uncertainties one needs to define the precise procedure for
evaluating the r.h.s.\ of Eq.\ (34b) at $\vec{r} = 0$.  It is assumed here
that the averaging is performed first over the angles between $\vec{r}$ and
$\vec{p}$ and then the value at the origin is computed.  The modified
expression for the $t\bar{t}$ production amplitude near threshold can be
presented in the following form (cf.\ Eq.\ (8))
\begin{equation}
A_{fi} \; = \; -4\pi\alpha\cdot e_t^2 \psi^+ \left[ a\tilde{G}(\vec{p},E) +
\frac{p}{m_t} \vec{b}\cdot \vec{\sigma}\cdot \tilde{F}(\vec{p},E) \right] \chi
\cdot \left(E - \frac{\vec{p}^2}{m_t} + i\Gamma_t \right) .
\end{equation}
In Eq.\ (35) the $a$ and $\vec{b}$ parameters are given by the relations (9)
and
\begin{equation}
p^2 \tilde{F}(\vec{p},E) \; = \; p_i \tilde{F}_i (\vec{p},E).
\end{equation}
One can easily recognize $\tilde{G}(\vec{p},E)$ as the Fourier transform of
the Schr\"{o}dinger Green's function
\begin{equation}
G_{E + i\Gamma_t}(\vec{r},\vec{r}\,^{\prime}) \; = \; \langle \vec{r}\,
|(\hat{H} - E - i\Gamma_t)^{-1}|\vec{r}\,^{\prime} \rangle
\end{equation}
for the relative motion of the $t\bar{t}$ pair evaluated at
$\vec{r}\,^{\prime} = 0$, and $-i\tilde{F}_k(\vec{p},E)$ as the Fourier
transform of its space derivative
\begin{displaymath}
\left[ \frac{\partial}{\partial x^{\prime}_k} G_{E + i\Gamma_t}
(\vec{r},\vec{r}\,^{\prime}) \right] \left|_{\vec{r}\,^{\prime} = 0} . \right.
\end{displaymath}
Indeed, returning (34a),(34b) to the coordinate space we see that the
functions
\begin{displaymath}
G(\vec{r},E) \; = \; G_{E+i\Gamma_t}(\vec{r},\vec{r}\,^{\prime} = 0) ,
\end{displaymath}
\begin{equation}
F_k(\vec{r},E) \; = \; i \left[ \frac{\partial}{\partial x^{\prime}_k}
G_{E+i\Gamma_t} (\vec{r},\vec{r}\,^{\prime}) \right]
\left|_{\vec{r}\,^{\prime} = 0} \right.
\end{equation}
satisfy the equations
\begin{displaymath}
(\hat{H} - E - i\Gamma_t)G(\vec{r},E) \; = \; \delta^{(3)}(\vec{r}),
\end{displaymath}

\begin{equation}
(\hat{H} - E - i\Gamma_t)F_k(\vec{r},E) \; = \; -i\frac{\partial}{\partial
x_k} \delta^{(3)} (\vec{r}) .
\end{equation}
Note that the functions $G(\tilde{G}), \, F_i(\tilde{F}_i)$ coincide with
those introduced in Ref.\ [23] for the description of the effects of the axial
current to the $t\bar{t}$ threshold production in the reaction (1).

One can expand the functions $\tilde{G}(\vec{p},E)$ and
$\tilde{F}_i(\vec{p},E)$ in terms of the solutions $\psi_{nlm}(\vec{r}), \,
\psi_{klm}(\vec{r})$ of the Schr\"{o}dinger equation with the Hamiltonian
$\hat{H}$ and their Fourier transforms,
\begin{equation}
\psi_{nlm(klm)}(\vec{p}) \; = \; \int d^3r e^{-i\vec{p}\cdot \vec{r}}
\psi_{nlm(klm)}(\vec{r})
\end{equation}
as
\begin{displaymath}
\tilde{G}(\vec{p},E) \; = \; \left\{ \sum^{\infty}_{n=1}
\frac{\psi_{noo}(\vec{p})\psi^*_{noo}(\vec{r} = 0)}{E_{no} - E - i\Gamma_t} +
\right.
\end{displaymath}

\begin{displaymath}
\left. + \int^{\infty}_0 \frac{dk}{2\pi} \frac{\psi_{koo}(\vec{p})
\psi^*_{koo}(\vec{r}=0)}{\frac{k^2}{m_t} - E - i\Gamma_t} \right\} ,
\hspace*{3cm} (a)
\end{displaymath}

\begin{displaymath}
-i \tilde{F}_i(\vec{p},E) \; = \; \sum^1_{m=-1} \frac{\partial}{\partial x_i}
\left\{ \sum^{\infty}_{n=2}
\frac{\psi_{n1m}(\vec{p})\psi^*_{n1m}(\vec{r})}{E_{n1} - E - i\Gamma_t} +
\right.
\end{displaymath}

\begin{equation}
\left. + \int^{\infty}_0 \frac{dk}{2\pi} \frac{\psi_{k1m}(\vec{p})
\psi^*_{k1m}(\vec{r})}{\frac{k^2}{m_t} - E - i\Gamma_t} \right\}
\left|_{\vec{r}=0} \right. .  \hspace*{2.75cm} (b)
\end{equation}
The reader is reminded that only $P$-wave states contribute to
$F_i(\vec{p},E)$.  Both expressions in Eq.\ (41) clearly exhibit two
contributions.  The first one corresponds to the production of the $t\bar{t}$
bound states with the binding energies $|E_{nl}|$.  The second one describes
the open top production with the energy $\frac{k^2}{m_t}$.  If one ignores
the running of the QCD coupling $\alpha_s$, then $E_{nl}$ depends only on $n$
(well known Coulomb degeneracy) and
\begin{displaymath}
E_{nl} \; = \; E_n \; = \; -\frac{p^2_s}{m_tn^2} ,
\end{displaymath}

\begin{equation}
p_s \; = \; \frac{2}{3} \alpha_s m_t .
\end{equation}
Expanding $\psi_{nlm(klm)}(\vec{r})$ over the radial wave functions
$R_{nl(kl)}(r)$
\begin{equation}
\Psi_{nlm(klm)}(\vec{r}) \; = \; R_{nl(kl)}(r) Y_{lm} \left( \frac{\vec{r}}{r}
\right)
\end{equation}
then repeating the anologous procedure for the Fourier transforms
\begin{equation}
\psi_{nlm(klm)}(\vec{p}) \; = \; \tilde{R}_{nl(kl)}(p) Y_{lm} \left(
\frac{\vec{p}}{p} \right)
\end{equation}
and making use of the relation
\begin{equation}
\sum_m Y_{1m} (\vec{n}) Y^*_{1m}(\vec{n}^{\prime}) \; = \; \frac{3}{4\pi}
(\vec{n}\cdot \vec{n}^{\prime})
\end{equation}
one can perform a summation over $m$ in Eq.\ (41b).  Then equations (41) can
be written as follows
\begin{displaymath}
\tilde{G}(\vec{p},E) \; = \; \frac{1}{4\pi} \left\{ \sum^{\infty}_{n=1}
\frac{\tilde{R}_{n0}(p)R_{n0}(r=0)}{E_{n0} - E - i\Gamma_t} +  \right.
\end{displaymath}

\begin{displaymath}
\left. + \int^{\infty}_0 \frac{dk}{2\pi} \frac{\tilde{R}_{k0}(p)
R_{k0}(r=0)}{\frac{k^2}{m_t} - E - i\Gamma_t} \right\}  \hspace*{3cm} (a)
\end{displaymath}

\begin{displaymath}
-i \tilde{F}_i(\vec{p},E) \; = \; \frac{p_i}{p} \cdot \frac{3}{4\pi} \left\{
\sum^{\infty}_{n=2} \frac{\tilde{R}_{n1}(p)R_{n1}^{\prime}(r=0)}{E_{n1} - E -
i\Gamma_t} +  \right.
\end{displaymath}

\begin{equation}
\left. + \int^{\infty}_0 \frac{dk}{2\pi} \frac{\tilde{R}_{k1}(p)
R_{k1}^{\prime}(r=0)}{\frac{k^2}{m_t} - E - i\Gamma_t} \right\}  \hspace*{3cm}
(b)
\end{equation}
The functions $\tilde{R}_{nl(kl)}(p)$ can be expressed in terms of the radial
wave-functions $R^{(0)}_{pl}(r)$ for the free motion as
\begin{equation}
\tilde{R}_{nl(kl)}(p) \; = \; \frac{(-i)^l}{2p} \int^{\infty}_0 r^2 dr \left\{
R_{nl(kl)}(r)R^{(0)}_{pl}(r) \right\} .
\end{equation}
Then
\begin{displaymath}
\tilde{R}_{n0(k0)}(p) \; = \; \int^{\infty}_0 r^2 dr \left( R_{n0(k0)}(r)\cdot
\frac{{\rm sin}pr}{pr} \right) , \hspace*{3cm} (a)
\end{displaymath}

\begin{equation}
\tilde{R}_{n1(k1)}(p) \; = \; \frac{i}{p} \int^{\infty}_0 r^2 dr \left(
R_{n1(k1)}(r) \cdot \frac{d}{dr} \left( \frac{{\rm sin}pr}{pr} \right)
\right).  \hspace*{2cm} (b)
\end{equation}

\vspace*{1cm}
\noindent 4.  TOP SPIN PARAMETERS  \\
\indent The amplitude $A_{fi}$ allows one to derive the threshold expressions
for both the unpolarized cross section $d\sigma^{(0)}_{\gamma\gamma}$ for the
top production and the $t$-polarization vector.  Performing summation over
the quark spins and using Eq. (20) for phase space element we arrive at the
expression for the unpolarized cross section
\begin{equation}
d\sigma^{(0)}_{\gamma\gamma} \; = \;
\frac{\alpha^2R_{\gamma\gamma}}{4\pi m_t^4} \cdot \Gamma_t d^3p \left
|a|^2|\tilde{G}(\vec{p},E)|^2  \right
\end{equation}
with $|a|^2$ given by Eq. (10) and $\tilde{G}$ given by Eq. (34a).
The reader is reminded that the total cross-section is obtained through the
equation$^{[20,25,26]}$
\begin{equation}
\int \frac{d^3p}{(2\pi)^3} \Gamma_t |\tilde{G}(\vec{p},E)|^2 \; = \; Im
G_{E+i\Gamma_t} (\vec{r} = 0, \vec{r}\,^{\prime} = 0) .
\end{equation}

In addition to the Coulomb
ladder exchanges one can also include the order  $\alpha_s$ hard gluon
corrections to the cross-section.  They give an additional factor
\begin{equation}
1 + \delta_h \; = \; 1 - \frac{4}{3} \frac{\alpha_s(m_t^2)}{\pi} \left( 5 -
\frac{\pi^2}{4} \right) .
\end{equation}
As a matter of fact, this result easily comes from the one-loop
calculations$^{[27]}$ of the radiative corrections to the pure QED process
$e^+e^- \rightarrow \gamma\gamma$ (with $\alpha$ substituted by $\frac{4}{3}
\alpha_s(m^2_t)$).\footnote{For a general analysis of the $O(\alpha_s)$ QCD
corrections to the cross section of the process (2) see Ref.\ [16].}
Moreover, the authors of Ref.\ [27] were the first to suggest the separation
of the exact (all-order) Coulomb contribution and the first order hard
correction into two multiplicative factors.  This separation is
rooted in the difference between the Coulomb phenomena occurring at the
characteristic distances $d_c \sim \frac{1}{p}$ and the hard correction
related to the dynamics at much smaller distances $d_h \sim \frac{1}{m_t}$.
The Coulomb forces cannot affect the short distance result and both
contributions factorize.

An attempt to keep under full control all order-$\alpha_s$ corrections to the
cross-section for the production of unstable particle may face some potential
problems arising from accounting for the interferences between the radiative
decays of these particles.  However, in the case of total cross sections the
radiative interference effects are shown$^{[28,29]}$ to be of $O(\alpha_s
\frac{\Gamma_t}{m_t})$ or better.\footnote{We do not agree with the statement
of Ref.\ [23] that an account of the gluon exchanges between $t$ and $\bar{b}$
($\bar{t}$ and $b$), and between $b$ and $\bar{b}$ could affect sizeably the
threshold results for the forward-backward asymmetry $\Delta_{FB}$.}
Therefore all the order-$\alpha_s$ effects in the cross section
$\sigma^{(0)}_{\gamma\gamma}$ are correctly included by taking account of the
running of $\alpha_s$ and the hard correction (51) assuming that $\Gamma_t$ is
the physical width incorporating the radiative top decays.$^{[28]}$

Let us return to the $t$-quark polarization.  With account of the expression
(20) for the phase space volume the threshold cross section
$d\sigma^{(\vec{\zeta}^{\prime})}_{\gamma\gamma}$ for polarized top production
can be written as
\begin{displaymath}
d\sigma^{(\vec{\zeta}^{\prime})}_{\gamma\gamma} \; = \;
\frac{\alpha^2R_{\gamma\gamma}}{8\pi m_t^4} \cdot \Gamma_t d^3p \left[
|a|^2|\tilde{G}(\vec{p},E)|^2 +  \right.
\end{displaymath}

\begin{equation}
\left.  + \frac{2p}{m_t} Re (a^*(\vec{b}\cdot \vec{\zeta}^{\prime}) \cdot
\tilde{G}^*(\vec{p},E) \tilde{F}(\vec{p},E)) \right]
\end{equation}
with $R_{\gamma\gamma}, |a|^2$ and $a^*(\vec{b}\cdot \vec{\zeta}^{\prime})$
given by Eqs.\ (5),(10) and (12),(14) respectively (cf.\ Eq.\ (4)).  An
expression for the $t$-quark spin vector $\zeta$ is then
\begin{displaymath}
\vec{\zeta} \; = \; 2 \frac{p}{m_t} Re(\vec{\zeta}_0\cdot D(\vec{p},E)) ,
\hspace*{3cm} (a)
\end{displaymath}

\begin{equation}
D(\vec{p},E) \; = \; \frac{\tilde{F}(\vec{p},E)}{\tilde{G}(p,E)} ,
\hspace*{4cm}  (b)
\end{equation}
where $\vec{\zeta}_0$ is given by Eq. (16).

Thus the top polarization could be strongly modified by the threshold effects
and its measurement in collisions of polarized photons would provide a
challenging opportunity to probe experimentally the new characteristics of the
interquark potential, $Re(\tilde{G}^*\cdot \tilde{F})$ and
$Im(\tilde{G}^*\tilde{F})$.
The latter quantity looks especially intriguing since it cannot be measured by
the other methods which are now under consideration, see e.g.\ [23].  To some
extent, however, this should be taken with a pinch of salt.  The point is that
the correlation $Im(a^*\vec{b}\vec{\zeta}^{\prime})$ is constructed from the
linear polarizations of the incoming photons (see section 2).  But at PLC one
can expect the high degree only for the circular polarization (about 95\% for
polarized electron beams.\footnote{VAK is indebted to D.\ Borden for the
discussion of the polarization properties of the beams at PLC.}). The degree
of linear polarization will not be as high as the circular one (probably
around 30\%).  Moreover, the monochromaticity of the beams is highest just in
the case of circular polarization and as is well known, the monochromaticity
is one of the main concerns in the discussions of the threshold $t\bar{t}$
production.  However, we do not expect that the top polarization gets very
much diluted because of the beam energy spread and, in principle, better
monochromaticity could be achieved at the cost of luminosity.  Once more, one
has to sacrifice more luminosity in the linear case than in the circular case.
It would be rather premature (and for these authors also quite inappropriate)
to make some more definite statements about the prospects of exploiting the
advantages of the linear polarization as well as about the effects of the
energy resolution for the beams that might be achievable ten years from now.

In the case of the \lq\lq PLC friendly" circularly polarized photons
the basic vector $\vec{\zeta}_0$ is real,so the
threshold phenomena do not affect the orientation of the top spin which is
directed along  $\vec{\zeta}_0$.  But they may strongly
modify the degree of polarization.  The corresponding rescaling factor
\begin{equation}
\lambda \; = \; \frac{|\vec{\zeta}|}{|\vec{\zeta}_0|} \; = \;
2\frac{p}{m_t} Re D(\vec{p},E)
\end{equation}
($\lambda = \lambda_B = 2\frac{p}{m_t}$ in the Born case, cf.\ Eq.\ (16) with
Eq.\ (53)) is directly related to the quantity $\frac{T_1}{T_0}$ describing
the
$t$-quark forward-backward asymmetry effects in $e^+e^-$ collisions$^{[23]}$
\begin{equation}
\lambda \; = \; C\frac{T_1}{T_0}
\end{equation}
where $C$ is a known $p$-independent factor.  Numerically, $C \simeq 2\left( 1
- \frac{m^2_z}{s} \right)$.

Performing in Eq.\ (52) the $dp$ integration (keeping the production angle
$\theta$ fixed) we arrive at the following expression for the averaged
polarization vector
\begin{displaymath}
\langle \vec{\zeta}\rangle_p \; = \; \frac{2}{m_t} Re(\vec{\zeta}_0 \cdot I) ,
\end{displaymath}

\begin{equation}
I \; = \; \frac{\int^{p_m}_0 p^3 dp \tilde{G}^*(\vec{p},E)
\tilde{F}(\vec{p},E)}{\int p^2 dp |\tilde{G}(\vec{p},E)|^2} .
\end{equation}
Recall that the integrand in the denominator is calculated through the
Eq.(50).
Note that we introduced the upper limit of integration $p_m$ in the numerator
since the integral formally logarithmically diverges in the region of large
momenta.  This is related to a known problem of the non-relativistic treatment
of the $P$-wave production which proves to be not fully
self-consistent$^{[22,23]}$.  The upper limit arises from the kinematical
condition relating $p_m$ to the case when one quark is on-mass-shell and the
other one has the mass equal to a minimal invariant mass of its decay
products$^{[22]}$.  At $m_b = 0$ this gives
\begin{equation}
p_m \; = \; \frac{1}{4m_t} \sqrt{(9m^2_t-m_W^2)(m_t^2-m^2_W)} .
\end{equation}
For the case at hand ($t$-quark polarization given by Eq.\ (56)) the
dependence on $p_m$ is very weak.
There are two reasons for the weak influence of the high momentum
contribution.  First, the numerical factor accompanying $ln(p^2_m)$ is rather
small (order $\frac{\Gamma_t}{m_t})$.  Second, such a logarithmic term arises
only in the Born piece of the function $\tilde{G}^*\tilde{F}$ which plays a
minor role in the threshold region.

\vspace*{1cm}
\noindent 5.  DISCUSSION \\
\indent In the threshold region the ratio $D$ (see Eq.\ (53)) exhibits a quite
intricate structure which is strongly  dependent on the top quark mass and the
precise value of $\alpha_s$.  To elucidate some specific consequences of the
Coulomb physics let us consider first what happens in the zero width
approximation above the threshold
\begin{equation}
\Gamma_t \rightarrow 0 , \hspace*{1cm} E \rightarrow \frac{p^2}{m_t} ,
\hspace*{1cm} p \rightarrow \beta m_t
\end{equation}
where the integral cross section proves to be strongly modified by the final
state interactions
\begin{equation}
\frac{d\sigma}{d\sigma^{(B)}} \; = \; \frac{|R_{k0}(0)|^2}{4m_tE} \; = \;
|\Psi_{(s)}(0)|^2 .
\end{equation}
Here $|\Psi_{(s)}(0)|^2$ is a QCD analogue of the famous Sommerfeld-Sakharov
factor$^{[30]}$ ($\Psi_{(s)}(0)$ is the wave function at the origin for a
colour singlet $S$-wave state).  For pure Coulomb interaction (fixed
$\alpha_s$)
\begin{equation}
|\Psi_{(s)}(0)|^2 \; = \; \frac{Z}{1 - e^{-Z}} ,
\end{equation}
\begin{equation}
Z \; = \; \frac{2\pi p_s}{p} .
\end{equation}
To evaluate the factor $D(\vec{p},E)$ in the stable quark limit it is
convenient to use the representations (34) and to multiply both the numerator
and the denominator on the r.h.s.\ of Eq.\ (53b) by the Born propagator
function $\left( \frac{p^2}{m_t} - E - i\delta \right)^{-1}$.  Then one can
make use of the limiting relation
\begin{equation}
\left( \frac{p^2}{m_t} - E
 \mp i\delta \right) (\hat{H} - E \mp i\delta)^{-1} |\vec{p}\rangle
\left|_{\stackrel{E \rightarrow \frac{p^2}{m_t}}{\delta \rightarrow +0}} \; =
\; |\vec{p}_{\pm} \rangle \right.
\end{equation}
which follows directly from the Lippman-Schwinger equation for the incoming
$(|\vec{p}_+\rangle)$ and outgoing $(|\vec{p}_-\rangle)$ states
\begin{equation}
|\vec{p}_{\pm}\rangle \; = \; |\vec{p}\rangle +
 (\frac{1}{\frac{p^2}{m_t} -
\hat{H}^0 \pm i\delta} V |\vec{p}_{\pm}\rangle )\left|_{\delta \rightarrow +0}
\right.
\end{equation}

Finally, we arrive at the limiting expression for the spin-analyzing function
\begin{equation}
D^*(\vec{p},E)|_{E \rightarrow \frac{p^2}{m_t}, \Gamma_t \rightarrow 0} \; =
\;
\frac{\vec{p}\langle\vec{r}|\hat{\vec{p}}|\vec{p}_-\rangle}{p^2\langle
\vec{r}|\vec{p}_-\rangle} \left|_{\vec{r}=0} \; = \; -\frac{i}{p^2}
\vec{p}\cdot \vec{\nabla} \ell n \psi^-_{\vec{p}} (\vec{r})\right|_{\vec{r}=0}
\end{equation}
with
\begin{equation}
\psi^-_{\vec{p}}(\vec{r}) \; = \; \langle \vec{r}|\vec{p}_-\rangle .
\end{equation}
Making use of partial expantion for $\psi^-_{\vec{p}}(\vec{r})$ we have
from (64)
\begin{equation}
D^*(\vec{p},E)|_{E \rightarrow \frac{p^2}{m_t}, \Gamma_t \rightarrow 0} \; =
\;
e^{-i (\delta_1-\delta_0)} \frac{1}{p} \frac
{R_{p1}^{\prime}(r=0)}{R_{p0}(r=0)} ,
\end{equation}
where $\delta_l$ is a scattering phase for the angular momentum $l$.

Using for the case of fixed $\alpha_s$ the known representation for the
function $\psi^-_{\vec{p}}(r)$ $^{[31]}$
\begin{displaymath}
\psi^-_{\vec{p}}(\vec{r}) \; = \; e^{\frac{Z}{4}} \Gamma \left( 1 +
i\frac{Z}{2\pi} \right) e^{i\vec{p}\cdot\vec{r}} \cdot
\end{displaymath}

\begin{equation}
\cdot F\left( -i\frac{Z}{2\pi}, 1, -i(pr + \vec{p}\cdot\vec{r})\right)
\end{equation}

one obtains
\begin{equation}
D\left|_{\stackrel{\Gamma_t = 0,}{p^2 = mE}} \; = \; \left( 1 -
i\frac{Z}{2\pi} \right) \right.
\end{equation}
which leads to the striking result
\begin{equation}
\vec{\zeta}|_{\Gamma_t = 0} \; = \; \vec{\zeta}_B + \frac{4}{3} \alpha_s
Im(\vec{\zeta}_0)
\end{equation}
(see Eqs.\ (13),(16),(53)).

Therefore, in this limiting case the polarization of the top quark can be
affected by the final state QCD interaction only if the $t\bar{t}$ pair is
produced in collisions of the linearly polarized photons.  The nontrivial fact
that $Re D$ exactly equals 1 is a specific prediction for a pure Coulomb
potential.  This is not valid if one takes into account the running of the QCD
coupling.  At the same time, the effects related to $Im(\vec{\zeta}_0)$ are
the direct consequence of the interquark QCD interaction.
This demonstrates the high sensitivity of the
factor $D(\vec{p},E)$ to the behaviour of the QCD potential.
There are no reasons to expect that the
polarization asymmetries  in open top production get very much diluted by the
effects of energy distribution in the incoming beams.  Therefore, the studies
of the top polarization above threshold promise to be quite informative.

Let us include in the consideration the width of the top keeping $\alpha_s$
fixed.  Using Meixner representation for $G_E(\vec{r},\vec{r}^{\prime}=0)$
[32] we obtain
\begin{equation}
\tilde{G}(\vec{p},E) \; = \; \frac{m_t}{(\kappa^2+p^2)} [1+4p_s\kappa \cdot
\int^{1}_0 \frac{dx x^{-\frac{p_s}{\kappa}}}{\kappa^2(1+x)^2+p^2(1-x)^2}],
\end{equation}

\begin{equation}
D(\vec{p},E) \; = \; 2 + {(\kappa^2+p^2)} \frac{d}{dp^2}
\ell n \tilde{G}(\vec{p},E),
\end{equation}
where $\kappa = \sqrt{-m_t(E+i\Gamma_t)}$.  For
$\Gamma_t > \frac{m_t \alpha_s^2}{2 \sqrt{3}}$ representation (70) is
valid for all real values of $E$; in the opposite case it can be used in the
region
\begin{equation}
E^2+\Gamma^2_t > \frac{2}{9} m_t \alpha_s^2(\sqrt{E^2+\Gamma^2_t}-E).
\end{equation}
For values of $E$ outside this region $D(\vec{p},E)$ can be obtained
by analytical continuation.

It is clear from (70),(71), that $D(\vec{p},E) \simeq 1$ at $p \gg p_s$.  This
is in accord with  the region of applicability of Born approximation.
For running $\alpha_s$ it can be written as
\begin{equation}
\frac{p}{m_t} \gg \alpha_s(p)
\end{equation}
In the region (73) the influence of the final state interaction on the top
polarization is negligible.  In the opposite case of small $p$ one can
obtain from (70),(71)
\begin{equation}
D(\vec{0},E) \; = \; 8 \frac{\int^1_0 \frac{dxx^{1-\frac{p_s}{\kappa}}
(1-x)}{(1+x)^5}}{\int^1_0 \frac{dxx^{-\frac{p_s}{\kappa}} (1-x)}{(1+x)^3}}.
\end{equation}
This shows that outside the region (73) $D(\vec{p},E)$ strongly
depends on $\alpha_s$.  This conclusion is confirmed by the numerical
calculations for the realistic case of a running $\alpha_s$.
We have performed the calculations by a method similar to that of
Ref.[21,23,25],
using the same parameters as in [25].

The momentum dependence of $ReD$ and $ImD$ for $m_t = 150 GeV$ and $E = 0$
is shown in Fig.\ 3.   It is seen that the quantity
$D$ in the region $p < 20 GeV$ is strongly sensitive to value of $\alpha_s$.
For larger values of $p$ the sensitivity to $\alpha_s$ vanishes, as one would
expect.

The dependence of $D$ on $\alpha_s$ becomes weaker  for heavier quarks.  This
is illustrated by Fig.\ 4  where $ReD(\vec{p},0)$ and
$ImD(\vec{p},0)$ are presented for $m_t = 200 GeV$.  Weak  sensitivity of
$D$ to $\alpha_s$ at large $m_t$ as well as the smooth dependence on $p$ is
the
consequence of a steep growth of the width $\Gamma_t$ with increasing of the
mass $m_t$.  Because of this growth the influence of the final state
interaction decreases with increasing mass so the value of $D$ tends to
unity.

Let us now come to the region below threshold.  Here the $t$-quark
polarization is dominated by the interference between S and P-wave states.
The overlap of S-wave and P-wave states
rapidly increases with increasing top mass because of the steep growth of the
top quark width $\Gamma_t$.  Recall that for a pure Coulomb potential the
levels (2S,3S,...) and (2P,3P,...) are degenerate.  In QCD, because of the
running of $\alpha_s$ the P-wave levels are located slightly below the
corresponding S-wave counterparts.

For the purposes of illustration we shall restrict ourselves to the effects of
the interference between the ground state and the 2P level.  Recall that the
characteristic distances in the $t\bar{t}$ system in the region of the lowest
Coulomb bound states are of order of the Bohr radius
\begin{equation}
a_o \sim (m_t \alpha_s)^{-1}
\end{equation}
and that for $m_t \gapproxeq$ 140 GeV the level spacings and the width
$\Gamma_t$ have the same order of magnitude.$^{[5]}$  Evaluating the factors
$\tilde{R}_{21}(p), \, \tilde{R}_{10}(p)$ and $R_{10}(r = 0), \,
R_{21}^{\prime}(r =0)$ (see Eqs.\ (46),(48)) at these distances one can
estimate
\begin{displaymath}
\int p^3 dp \tilde{G}^*\tilde{F} \; \sim \; \frac{1}{(a_0)^4}
\frac{1}{(E_1-E+i\Gamma_t)} \frac{1}{(E_2-E-i\Gamma_t)}
\end{displaymath}

\begin{equation}
\int p^2 dp |\tilde{G}|^2 \; \sim \; \frac{1}{(a_0)^3} \left|
\frac{1}{E_1-E-i\Gamma_t} \right|^2 + ...
\end{equation}
Then it is easy to show that for relatively heavy top when
\begin{equation}
\Gamma_t > Ry
\end{equation}
(the Rydberg $Ry = \frac{4}{9} m_t\alpha^2_s)$ the degree of top polarization
$|{\langle \vec{\zeta}\rangle}_p| \sim O(\alpha_s)$ and its dependence on $E$
is rather smooth.    At
fixed top width $\langle\lambda_p\rangle$ decreases with increasing $\alpha_s$
because of the growth of the level spacing.  This is in accord with the
numerical results of Ref.\ [23] for $\Delta_{FB} \sim Re I$.  Thus one faces
once more a clear manifestation of the final state interaction in top
polarization effects.  These would not be essentially smeared out after
folding the corresponding cross-sections with the luminosity distributions in
the incoming beams.

Note that in the narrow width case
\begin{equation}
\Gamma_t \ll Ry
\end{equation}
the top polarization below threshold  can be a very steep function of
momentum $p$.  This is illustrated by Fig.\ 5.   The sharp dependence of
$D$ on $p$ can be understood using representation (41 a,b) for
$\tilde{G}(\vec{p},E)$ and $\tilde{F}(\vec{p},E)$.  If the energy $E$
is close to the $2S$ level, one can keep only contributions of $2P$ and $2S$
resonances in the functions $\tilde{F}$ and $\tilde{G}$ correspondingly.
Then we have
\begin{equation}
D(\vec{p},E) \simeq \frac{E-E_{20}+i\Gamma_t}{E-E_{21}+i\Gamma_t} \frac{
R_{21}^{\prime}(0)}{p R_{20}(0)} \frac{
3i R_{21}(p)}{R_{20}(p)}.
\end{equation}
We recall that the representation (79) can not be used literally for all
values
of $p$, because the radial wave function $R_{20}(p)$ according to the
\lq\lq oscillation theorem" passes through zero at some point $p = p_0$.
In the vicinity of $p_0$ one needs to take into account another
contributions to $\tilde{G}(\vec{p},E)$.  But nevertheless the
representation (79) provides a quantitative explanation of Fig.\ 5.

Increasing the top mass makes the dependence of $D$ on momentum
smooth even for the energy $E$ in the region of the $2S$ level due to a steep
growth of $\Gamma_t$.  It is seen from Fig.\ 6 which differs only from Fig.\ 5
by the choice $m_t$ = 150 GeV.  In the narrow width case (see Eq.\ (78)) the
average over $p$ polarization
will suffer smearing induced by the
energy spread in the beams.  To demonstrate this let us note that $Re I$
changes sign when $E$ crosses the poles $E_1, E_2$ and its magnitude reaches
the maximal value midway between these resonances.  Thus, $Re I$ oscillates
with $E$ increasing.  $Im I$ does not oscillate but its magnitude is always
small.  If the beam energy
spread $\sigma$ is much larger than $Ry$, then folding with the luminosity
distribution essentially reduces the degree of polarization
\begin{equation}
|{\langle \vec{\zeta}\rangle}_p| \; \sim \; O\left(
\alpha_s \frac{\Gamma_t}{Ry} \right) .
\end{equation}

\vspace*{1cm}
\noindent 6.  CONCLUSION \\
\indent We have studied the influence of the spectacular threshold effects on
the polarization of top quarks produced in the collisions of polarized
photons.  Without loss of generality one may consider some other parity
violating angular asymmetries in the distribution of the secondary particles.
For relatively heavy top quark $(m_t \gapproxeq$ 130 GeV) the decay width
$\Gamma_t$ provides an infrared cut-off for the strong forces between quarks
and antiquarks and perturbative QCD suffices to treat the threshold phenomena.

We have demonstrated that the final state interaction between $t$-quarks
induces two major modifications of the Born result for the top polarization
above threshold.  First, in the collisions of linearly polarized photons order
$\alpha_s$ (T-odd) effects may appear which would allow one to measure the
relative phase of the low energy S-wave and P-wave scattering of $t$-quarks.
Second, the degree of polarization arising in collisions of circularly
polarized photons can be seriously affected because of the running of the QCD
coupling.  For relatively heavy top quark its polarization below threshold is
quite sizeable, order $\alpha_s$.  The resulting polarization effects are
expected to survive the averaging over the luminosity distributions in the
incoming beams.  Optimistically, we anticipate that these polarization
measurements would open new prospects in studying the interquark dynamics.  In
combination with the threshold measurements of the cross-sections these
polarization results may provide us with an efficient systematic cross-check
on the accuracy with which the values for $m_t$ and $\alpha_s$ will have been
determined.

\newpage

\noindent Figure Captions

\begin{itemize}
\item[Fig.\ 1] Schematic description of the ladder diagram involving the
exchange of any number of Coulombic gluons.

\item[Fig.\ 2] The equation which is satisfied by the function $R(\vec{p})$
in the ladder approximation.

\item[Fig.\ 3] Dependence of $Re (\vec{p},0)$ (a) and $ImD(\vec{p},0)$ (b) on
$p$ at $m_t$ = 150 GeV for $\alpha_s(m_z)$ = 0.10, 0.12 and 0.14.

\item[Fig.\ 4] The same as Fig.\ 3 at $m_t$ = 200 GeV.

\item[Fig.\ 5] Momentum dependence of $ReD(\vec{p},E)$ (a) and
$ImD(\vec{p},E)$ (b) at $m_t$ = 100 GeV for $\alpha_s(m_z)$ = 0.10, 0.12 and
0.14.  For each value of $\alpha_s(m_z)$ the energy $E$ is fixed at the
position of the 2S level.

\item[Fig.\ 6] The same as Fig.\ 5 at $m_t$ = 150 GeV.
\end{itemize}

\newpage

\noindent REFERENCES
\begin{enumerate}
\item P.D.\ Grannis, Summary of the 9th $\bar{p}-p$ Workshop, Tsukuba, Japan,
Oct.\ 18-22 1993.
\item R.\ Rolandi, in Proc.\ XXVI Int.\ Conf.\ on High Energy Physics, Dallas
1992, Ed.\ J.R.\ Sanford, AIP Conf.\ Proc., V.\ 1, p.\ 56.
\item P.\ Langacker, Univ.\ of Pennsylvania preprint UPR-0555-T, March 1993.
\item  J.H.\ K\"{u}hn, Acta Phys.\ Aust.\ Suppl.\ 24 (1982) 203.
\item I.I.\ Bigi et al., Phys.\ Lett.\ 181B (1986) 157.
\item For recent Standard Model update of the top quark width see M.\
Je\.{z}abek and J.H.\ K\"{u}hn, Phys.\ Rev.\ D (to be published).
\item J.H.\ K\"{u}hn and K.H.\ Streng, Nucl.\ Phys.\ B198 (1982) 71; \\
J.H.\ K\"{u}hn, Nucl.\ Phys.\ B237 (1984) 77.
\item T.\ Arens and L.M.\ Sehgal, Nucl.\ Phys.\ B393 (1993) 46.
\item  Proceedings, \lq\lq Physics and Experiments with $e^+e^-$ Linear
Colliders", Saariselk\"{a} 1991, R.\ Orava, P.\ Eerola and M.\ Nordberg eds.,
World Scientific, Singapore 1992.
\item Proceedings, $e^+e^-$ Collisions at 500 GeV: The Physics Potential,
Munich-Annecy-Hamburg 1991, P.M.\ Zerwas ed., DESY 92-123 A,B.
\item P.\ Zerwas, Talk at LC92, Workshop on $e^+e^-$ Linear Colliders,
Garmisch-Partenkirchen, Germany, 1992; DESY 93-001, Jan.\ 1993. \\
J.H.\ K\"{u}hn and P.M.\ Zerwas, in Advanced Series on Directions in High
Energy Physics, \lq\lq Heavy Flavours", eds.\ A.J.\ Buras and M.\ Lindner,
World Scientific, Singapore 1992, p.\ 434.
\item I.F.\ Ginzburg et al., Nucl.\ Inst.\ Meth.\ 205 (1983) 47; \\
Nucl.\ Inst.\ Meth.\ 219 (1984) 5.
\item D.L.\ Borden, D.A.\ Bauer and D.O.\ Caldwell, SLAC-PUB-5715, Jan.\
1992;\\  Phys.\ Rev.\ D48 (1993) 4018.
\item  V.I.\ Telnov in [9], p.\ 739.
\item F.R.\ Arutyunian and V.A.\ Tumanian, Phys.\ Lett.\ 4 (1963) 176; \\
R.H.\ Milburn, Phys.\ Rev.\ Lett.\ 10 (1963) 75.
\item J.H.\ K\"{u}hn, E.\ Mirkes and J.\ Steegborn, Z.\ Phys.\ 57 (1993) 615.
\item I.I.\ Bigi, F.\ Gabbiani and V.A.\ Khoze, Nucl.\ Phys.\ B406 (1993) 3.
\item O.J.P.\ \'{E}boli et al., Phys.\ Rev.\ D47 (1993) 1889.
\item V.S.\ Fadin and V.A.\ Khoze, JETP Lett.\ 46 (1987) 525; \\
Sov.\ J.\ Nucl.\ Phys.\ 48 (1988) 309.
\item  V.S.\ Fadin and V.A.\ Khoze, Proc.\ 24th LNPI Winter School, Leningrad
1989, V.\ 1, p.\ 3.
\item M.J.\ Strassler and M.E.\ Peskin, Phys.\ Rev.\ D43 (1991) 1500.
\item V.S.\ Fadin and V.A.\ Khoze, Sov.\ J.\ Nucl.\ Phys.\ 53 (1991) 692; \\
I.I.\ Bigi, V.S.\ Fadin and V.A.\ Khoze, Nucl.\ Phys.\ B377 (1992) 461.
\item H.\ Murayama and Y.\ Sumino, Phys.\ Rev.\ D47 (1992) 82.
\item One can find the general Born-approximation formulae for polarization
effects in the process $\gamma\gamma \rightarrow e^+e^-$ in K.A.\ Ispiryan et
al., Sov.\ J.\ Nucl.\ Phys.\ 11 (1970) 712.
\item Y.\ Sumino et al., Phys.\ Rev.\ D47 (1992) 56; \\
Y.\ Sumino, Thesis, UT-655 (1993).
\item M.\ Je\.{z}abek, J.H.\ K\"{u}hn and T.\ Teubner, Z.\ Phys.\ C56 (1992)
653; \\
M.\ Je\.{z}abek and T.\ Teubner, Karlsruhe preprint, TTP 93-11.
\item I.\ Harris and L.M.\ Brown, Phys.\ Rev.\ 105 (1957) 1656.
\item V.S.\ Fadin, V.A.\ Khoze and A.D.\ Martin, Phys.\ Rev.\ D (to be
published); \\
Phys.\ Lett.\ 320B (1994) 141.
\item K.\ Melnikov and O.\ Yakovlev, Novosibirsk preprint INP 93-18 (1993).
\item A.\ Sommerfeld, \lq\lq Atombau and Spektrallinien", Bd.\ 2 (Vieweg,
Braunschweig, 1939); \\
A.D.\ Sakharov, JETP 18 (1948) 631.
\item L.D.\ Landau and E.M.\ Lifshitz, Quantum Mechanics (Non-Relativistic
Theory) Pergamon Press, 1977.
\item Meixner J., Math.Z. 36 (1933) 677.
\end{enumerate}

\end{document}